\documentclass[smallextended]{svjour3}       
\smartqed  
\usepackage{graphicx}
\usepackage{url, csquotes}
\usepackage{xcolor,colortbl}
\usepackage{todonotes}
\begin{document}

\title{Vindication, Virtue and Vitriol}
\subtitle{A study of online engagement and abuse toward British MPs during the COVID-19 Pandemic}


\author{Tracie Farrell \and
        Genevieve Gorrell \and Kalina Bontcheva 
}


\institute{T. Farrell \at
               Sheffield University \\
             \email{tracie.farrell@sheffield.ac.uk}           
           \and
           G. Gorrell \at
               Sheffield University \\
             \email{g.gorrell@sheffield.ac.uk}           
           \and
           K. Bontcheva \at
               Sheffield University \\
             \email{k.bontcheva@sheffield.ac.uk}           
}

\date{Received: date / Accepted: date}

\maketitle

\begin{abstract}
COVID-19 has given rise to malicious content online, including online abuse and hate toward British MPs. In order to understand and contextualise the level of abuse MPs receive, we consider how ministers use social media to communicate about the crisis, and the citizen engagement that this generates. The focus of the paper is on a large-scale, mixed methods study of abusive and antagonistic responses to UK politicians during the pandemic from early February to late May 2020. We find that pressing subjects such as financial concerns attract high levels of engagement, but not necessarily abusive dialogue. Rather, criticising authorities appears to attract higher levels of abuse. In particular, those who carry the flame for subjects like racism and inequality, may be accused of virtue signalling or receive higher abuse levels due to the topics they are required by their role to address.
This work contributes to the wider understanding of abusive language online, in particular that which is directed at public officials. 

\end{abstract}

\section{Introduction}
\label{intro}
Social media can offer a ``temperature check'' on which topics and issues are trending for certain cross-sections of the public, and how they feel about them~\cite{evans2011twitter}. This temperature has run high during the COVID-19 pandemic, with a number of \textbf{incendiary and misleading claims}~\cite{cinelli2020covid}, as well as \textbf{hateful and abusive content}~\cite{velasquez2020hate} appearing online. This content can interfere with both government and public responses to the pandemic. 

A recent survey of coronavirus conspiracy beliefs in England, for example, demonstrated that~\textbf{belief in conspiracy} was associated with lower compliance with government guidelines. Moreover, the authors found that 1 in 5 of their participants had a strong~\textbf{endorsement of conspiracy thinking}~\cite{freeman2020coronavirus}, indicating that this is not just a fringe issue.~\textbf{Online verbal abuse} contributes as well, being both cause and consequence of misinformation: the \textbf{quality of information and debate} is damaged as certain voices are silenced/driven out of the space,\footnote{\small{\url{https://www.bbc.co.uk/news/election-2019-50246969}}} and \textbf{escalation} leads to angry and aggressive expressions~\cite{sobieraj2011incivility}. Understanding the interplay between malicious online content and the public's relationships with authorities during a health crisis is necessary for an \textbf{effective response to the COVID-19 pandemic}. 

\subsection{Scope}
This work charts \textbf{Twitter abuse} in replies to UK MPs from \textbf{before the start of the pandemic in the UK, in early February, until late May 2020}, in order to plot the health of relationships of UK citizens with their elected representatives through the unprecedented challenges of the COVID-19 pandemic. We consider reactions to different individuals and members of different political parties, and how they interact with events relating to the virus. We review the \textbf{dominant hashtags} on Twitter as the country moves through different phases, as well as some \textbf{dominant conspiracy theories}. 

For these data, we show \textbf{trends in abuse levels}, for MPs overall as well as for particular individuals and for parties. We also compare \textbf{prevalence of conspiracy theories}, and contextualise them against other popular topics/concerns on Twitter. In addition to our quantitative analysis, we present an \textbf{in-depth qualitative analysis} on tweets receiving more than 20 abusive replies, that constitute 8\% or more of total replies that tweet received. 

We use a set of \textbf{qualitative codes} derived from the literature on how authorities make use of social media during crisis (and health crisis in particular). Referring to this as the \textbf{social media activities} of the MP who authored a tweet, we are able to label each tweet according to its most probable agenda (e.g. reaching out to constituents, communicating official information). This allows us to see the distribution of media activities across different parties and genders, for example. We then developed inductive codes for the \textbf{COVID-19 topic} and potential \textbf{controversial subject} of the tweet (e.g. communicating policy about COVID-19 or making criticisms of COVID-related policy and Brexit or Inequality, respectively), and noted any attached URLs or images for reference. We also listed the abusive words found in each reply to the tweet. Finally, we prepared an analysis of those labels by gender, party, sexual orientation and ethnicity.
In our analysis, we consider how \textbf{social dimensions of antagonistic political discourse} in the UK (ideology, political authority and affect), which have been visible in other recent key moments (such as Brexit and successive general elections), influence the civility of discourse during COVID-19.

\subsection{Contribution}
This study aims to answer the following research questions: 
\begin{enumerate}
    \item How has the context of COVID-19 impacted the typical patterns that have been observed in previous work about hateful and abusive language toward UK MPs?
    \item How do the social dimensions that have impacted political discourse on Brexit and successive general elections appear to impact how social media activities are perceived during the COVID-19 pandemic? 
    \item Which social media activities of UK MPs during the COVID-19 pandemic receive the most abusive replies? How can we contextualise these results?
   
\end{enumerate}

The contribution of this study is to understand both the content and context of \textbf{abusive and hateful communication, particularly toward governments and authorities during a health crisis}. Our focus, UK MPs, adds to a growing longitudinal body of work that analyses online abuse at many key moments in British politics from 2015 to the present~\cite{gorrell2018twits,gorrell2019race,binns2018and,ward2017turds}. 

In the sections below, we begin with a description of the current context and a brief summary of related work. We then outline our methodology in detail before progressing onto findings. Finally, we summarise and conclude our manuscript with some suggestions for future work.

\section{Context of COVID-19}
\label{sec:context}
The dangers and perceived risks of COVID-19 have fluctuated during the pandemic as a result of emerging knowledge. This feature of the pandemic creates an \textbf{environment of uncertainty and ambivalence} that feeds malicious content on social media. 

Early epidemiological studies of COVID-19 \cite{li2020early,yang2020clinical} implicated certain \textbf{risk factors}, such as age, gender and pre-existing conditions may impact transmission and severity of illness. As the pandemic progressed, researchers began to understand more about \textbf{asymptomatic transmission} \cite{bai2020presumed,nishiura2020estimation,day2020COVID}, discovering that there may be many more cases of COVID-19 than once realised. This led some researchers to suggest that the morbidity rate of COVID-19 is lower than initially presumed \cite{fauci2020COVID}, though this data is difficult to calculate\footnote{https://ourworldindata.org/mortality-risk-covid}. 

However, some communities were discovered to be at a \textit{greater} risk. Research on \textbf{health care professionals} who contracted COVID-19, for example, indicated that the most seriously ill individuals had \textbf{multiple exposures} to the virus, primarily at work~\cite{burrer2020characteristics}, as well as longer duration of (sometimes unprotected) exposure to the virus~\cite{heinzerling2020transmission}. As more information emerged about disproportionate cases of COVID-19 and poor health outcomes in  \textbf{Black and Asian communities} in England, researchers began to also investigate \textbf{early warning signs that some social or ethnic communities were more vulnerable than others}~\cite{khunti2020ethnicity,pareek2020ethnicity}. This too, has led to challenging debates about social welfare, racism and healthcare during the pandemic.

Tensions between competing political and social interests during times of uncertainly can lead to an increase in hateful and antagonistic discourse online. Valasquez \textit{et al} found that \textbf{malicious content of various kinds including misinformation, disinformation and hate speech are proliferating online about COVID-19}. Looking at clusters of online communities, the authors found that existing antagonistic groups appeared to mobilise COVID-19 to spread hate and malicious content even into mainstream communities~\cite{velasquez2020hate}. 


\section{Related Work}
\label{sec:RelatedWork}
Organisations use social media during crisis events to \textbf{correct rumours}, \textbf{prevent crisis escalation}, \textbf{provide facts} or information, \textbf{transmit proactiveness} toward resolving the situation, and to \textbf{communicate directly with members of the public} (without temporal or geographic constraints)~\cite{distaso2015managing}. Not using social media to address a crisis can incur reputational damage for the organisation~\cite{coombs2007protecting}.\footnote{https://www.pewresearch.org/internet/fact-sheet/social-media/}

Twitter and other forms of social media are popular tools used by organisations and governments to communicate with citizens during crisis events~\cite{petersen2017public}. The focus for the literature below is to briefly review how governments and authorities use such tools to communicate about health crises, particularly in the UK, and to explore how malicious content and abuse has been examined previously within this context. 

\subsection{Public Use of Social Media in a Health Crisis}
\label{Subsection:PubSocMed}
Before we address how politicians use social media in a health crisis, it is worth examining perspectives of the public and what they expect from politicians when emergencies like COVID-19 arise.  
Evidence indicates that the public expect a \textbf{swift, transparent response from the government to crisis}~\cite{boin2016politics,petersen2017public}.
The public may also wish to engage with the government on its response.  The greater the political interest of the user, the more likely they are to perceive and take advantage of the `\textbf{`connective affordances''} that social media provides for politicians and their constituents to engage~\cite{kalsnes2017social}. Third, and perhaps most importantly, during a health crisis, most citizens will be interested in \textbf{government advice and support}~\cite{veil2011work,llewellyn2020COVID}. 
For example, Vos and Buckner examined Tweets that were shared during H7N9 ``Bird Flu'' health crisis, and found that the majority of messages were about \textbf{``collective sense-making responses'' under conditions of uncertainty}, rather than ``efficacy responses'' offering specific advice or information that would help the public to respond appropriately~\cite{vos2016social}. A similar pattern was observed in response to the 2016 Zika virus outbreak, with individuals using social media to form a \textbf{personal risk assessment}~\cite{gui2017managing}. Llewellyn \textit{et al}~\cite{llewellyn2020COVID} found that the public seeks advice from experts and that the informal character of online communication can \textbf{interfere with the public's ability to form good opinions} about the expertise of individuals online, even public figures. If sense-making and risk-assessment are the top public tasks for which they seek information on social media, government messages that do not respond to this need may miss the mark. 


\subsection{Politicians' Social Media Use in a Health Crisis}
\label{subsec:PolSocMed}
In their analysis of political communication on social media, Stieglitz and Dang-Xuan~\cite{stieglitz2013social} show that politicians may use social media for \textbf{communication and persuasion}, to \textbf{``meet'' voters and engage them} in discussion, and also to \textbf{communicate policy or other important information} to their constituents. Political analysis of US congress members on Twitter shows that \textbf{self-promotion} is also an activity in which politicians engage, using the opportunity to share \textbf{personal information or stories}, and \textbf{present themselves and their platforms} in a good light~\cite{golbeck2010twitter}. However, this is not always true. Studies from the Swedish electoral context showed that Swedish politicians did not use Twitter to engage with voters, but rather to provide information to them~\cite{larsson2011tweets}. In the UK, because internal party campaigns are based on individual candidates, politicians in the UK share some media behaviours with their counterparts in the US, where individual voter appeal is critical to campaign success~\cite{lilleker2015social}. 

\subsubsection{Dimensions of Political Discourse in the UK}
The COVID-19 pandemic is a \textbf{novel political situation} in which ministers must respond to the crisis, while continuing to function in their roles. Though the situation may be new, the dialogue around COVID-19 is influenced by existing social and political dimensions of British political discourse. 

In their work documenting positions around the European Referendum, Andreouli \textit{et al} named three dimensions for understanding the emergent dialogue around Brexit~\cite{andreouli2019brexit}, that we feel may be useful here. These are \textbf{``political values, political authority, and the authority of affect''}. 

With regard to values, the authors reflect on how existing \textbf{ideological themes} impact how an issue is perceived and discussed, in particular, where classical dichotomies do not hold up. For example, while the left typically associates itself with anti-prejudice and tolerance, associating such qualities with voting to ``remain" is inconsistent with other leftist ideas to be anti-establishment. The authors argue that this tension creates a \textbf{``liminal hotspot''} where cosmopolitanism and critiques of globalisation intersect.  

We propose this same step, in light of the current crisis, to understand the dominant political and social themes that influence abuse toward UK MPS during COVID-19. We are already seeing evidence of potential areas of tension in the current pandemic, such as the \textbf{needs of older}\footnote{https://www.gov.uk/government/publications/coronavirus-COVID-19-support-for-care-homes/coronavirus-COVID-19-care-home-support-package} \textbf{and younger people} \footnote{https://www.ukyouth.org/wp-content/uploads/2020/04/UK-Youth-COVID-19-Impact-Report-External-Final-08.04.20.pdf}, the \textbf{reliance on science and perception of risk}~\cite{brzezinski2020belief,freeman2020coronavirus}, the division between the \textbf{wealthy and the poor}~\cite{walker2020report}, and the experiences of the \textbf{urban and rural}~\cite{iacobucci2020COVID,marston2020COVID}. 

Second, Andreouli \textit{et al}~\cite{andreouli2019brexit} discuss the notion of \textbf{political authority}, and how the sovereignty of the UK within the EU becomes a backdrop for discourse on immigration during the European Referendum. During the COVID-19 pandemic, the \textbf{sovereignty of local governments within the UK} has been a consistent feature of debate, whether it involved avoiding beauty spots in Wales during lockdown\footnote{https://www.bbc.co.uk/news/uk-wales-52614204}, comparing Scotland's success in handling the virus\footnote{https://www.bbc.co.uk/news/uk-scotland-53195166}, or the differential impact on the economy in Northern Ireland.\footnote{https://www.bbc.co.uk/news/uk-northern-ireland-53386976} 
Media reports during the peak of the outbreak also indicated resistance toward lockdown,\footnote{https://www.theguardian.com/uk-news/2020/may/14/police-vow-to-break-up-planned-anti-lockdown-protests-in-uk-cities}\footnote{https://www.telegraph.co.uk/news/2020/04/20/coronavirus-world-erupts-protest-against-lockdown-pictures/}, or wearing face-coverings\footnote{https://www.washingtonpost.com/world/europe/face-masks-coronavirus-uk/2020/07/14/d05dfb7c-c5d4-11ea-a825-8722004e4150\_story.html}\footnote{https://www.telegraph.co.uk/news/2020/04/06/britains-hubristic-scientific-advisers-wrong-public-should-wearing/}. As we move toward the next phases of the crisis, conflicts about personal agency and choice that may play out at individual or group levels in how the public respond to government guidance. 

Finally, Andreouli \textit{et al}~\cite{andreouli2019brexit} discuss the role of \textbf{affect} in political discourse. They demonstrate how \textbf{impassioned speech} has become its own kind of credibility, in which the narrative, rather than being factual, is important. In September 2019, parties signed a pledge to use moderate language after a series of heated and antagonistic debates in which Boris Johnson was criticised for framing the Brexit conversation as ``surrender'' to the European Union, and for invoking the name of British MP Jo Cox, who was murdered by a far-right extremist one week before the referendum.\footnote{https://www.theguardian.com/politics/2019/sep/30/westminster-parties-sign-  pledge-to-use-moderate-language\\ https://www.bbc.co.uk/news/uk-politics-49833804\\ https://www.bbc.co.uk/news/uk-37978582} Dawn Butler, who called the Prime Minister reckless in his decision to send employees back to work, was accused of using hyperbolic language in describing how such a policy amounted to sending people to work to catch the virus\footnote{https://www.kilburntimes.co.uk/news/coronavirus-dawn-butler-mp-calls-boris-johnson-reckless-1-6651422}. Due to her vocal support of \#blacklivesmatter, Butler has now had to shut down her offices in response to racism, death threats and other threats of violence.\footnote{https://www.bbc.co.uk/news/uk-england-london-53346803} 

One of the goals of this research is to analyse topics of discussion and responses within these three dimensions (see Section \ref{findings:dimensions}). This will allow us to contextualise the public's antagonistic responses to UK MPs on Twitter during COVID-19 thus far.

\subsubsection{Hate and Abuse of British MPs Online}
\label{subsec:HateBritishMPs}
As the conversation around online abuse develops, we need to \textbf{differentiate the hostility that arises from increased visibility and engagement, from that which is based on hate or hate-speech} toward a specific group of individuals or communities \cite{rossini2019toxic}.

In the UK, where legal frameworks tend to evolve, hate speech was defined through several legal statues, including The Public Order Act of 1986. \footnote{http://www.legislation.gov.uk/ukpga/1986/64/contents} Legal philosopher Jeremy Waldron~\cite{waldron2012harm} has. argued, however, that \textbf{hate speech is that which sends a message to undervalued groups of communities} that they are not welcome or wanted. It also \textbf{sends a message to other sympathisers} who may devalue or feel solidarity with that community. In particular, hate speech is associated with power~\cite{perry2001name}. 

Governments and politicians communicate hateful messages, for example, through language about Romani people in Europe~\cite{scicluna2007anti} or Mexican and other Latin immigrants in the United States~\cite{cervantes1995hate,kirk2017dark}. Governments have contributed to hate through politicising tribal identity in sub-Saharan countries like Kenya and Rwanda~\cite{somerville2011violence,faustin2016preventing}, and how they shape debates about free speech~\cite{pettersson2019freedom} or migration~\cite{vazquez2019hate} more generally. \textbf{Politicians, therefore, can both be the targets of hate (as members of protected groups) and the perpetrators (as public authorities whose words matter)}. Governments can also antagonise the public. In the context of COVID-19, the ways in which politicians are communicating about potentially volatile issues, such as re-opening the economy or avoiding social protest, add to the overall ``health'' of the discourse, or diminish it. 

There is a considerable amount of historical data on the prevalence of online abuse directed at British MPs, particularly on Twitter. Previous work~\cite{gorrell2018twits,gorrell2019race,binns2018and,ward2017turds} has shown rising levels of hostility towards UK politicians on Twitter, particularly in the context of \textbf{divisive issues, such as Brexit or inequality}. Partisan operators have been implicated in fanning the flames with malicious content, such as misinformation or troll accounts~\cite{gorrell2019partisanship}. 

In their 2017 and 2020 papers, Ward and McLoughlin examined online abuse received by British MPs from November 14, 2016 - January 28, 2017 \cite{mcloughlin2017turds,ward2020turds}. The major findings from this work were that the amount of \textbf{online hate (rather than language that can be described as ``abusive'' or uncivil) is relatively low} and, as such, men receive more online abuse than women. The authors also showed that \textbf{increased name recognition and popularity} have a positive relationship with levels of abuse. Crucially, however, the authors note that \textbf{women and those from a minority background} are more likely to receive abusive replies that can be classified as hate speech. Abuse toward specific parties was difficult to distinguish, as levels of abuse may be influenced by one party member who attracts a significant proportion of abuse. When controlling for this, the authors found that less visible MPs had a very small percentage of hate and abuse. This means that women MPs with visibility disproportionately attract hate speech, as do men with visibility other forms of abusive language. This work prompted questions about \textbf{what visibility means for people of different genders and backgrounds}. Southern and Harmer \cite{southern2019twitter} conducted a deeper content analysis on tweets received by MPs during a period and found that while men received more incivility in terms of numbers of replies, women were more likely to receive an uncivil reply. Women were more likely to be stereotyped by identity (men by party) and to be questioned in their position as an MP. 

Gorrell \textit{et al}~\cite{gorrell2020politicians} extended this work to define four visibility factors that appear to influence the amount of abuse the UK MPs receive online:
\begin{itemize}
    \item Prominence: Individuals in the public eye will receive more abuse;
    \item Event surges: Events leads to spikes in abuse (such as participation in an event, or political activities);
    \item Engagement: Expressing strong opinions on social media can result in more personal abuse;
    \item Identity: Gender, ethnicity and other personal factors impact which opinions one is allowed to hold and express without receiving abuse. 
    \end{itemize}
    
Gorrell \textit{et al} also note that the impacts or consequences of abusive language are not manifesting in the same ways for male and female MPs, or MPs with intersectional identities of race and gender. Where some abuse is distressing, other abuse is personal, threatening and limits women's participation in the public office~\cite{gorrell2019race,delisle2019large,pew2017}. 

From this review, a picture emerges of the \textbf{precipitating activities, mediating factors and dimensions of online abuse} toward UK MPs during COVID-19, which can be interrogated through our large-scale study. 

\section{Methodology}
\label{sec:Methodology}
In this work we apply a combination of computational and social science methods to evaluate abuse toward UK MPs on Twitter. We utilise a large tweet collection on which natural language processing analysis has been performed in order to identify abusive language. This methodology is presented in detail by Gorrell \textit{et al}~\cite{gorrell2019race} and summarised here. We then follow Braun and Clarke's~\cite{braun2012thematic} process of thematic analysis on a subset of tweets that received 8\% or more of abusive replies, in which at least 20 abusive replies were received. This analysis is described in more detail below. 

\subsection{Corpus}
\label{subsec:corpus}

The corpus was created by collecting tweets in real-time using Twitter's streaming API. We used the API to follow the accounts of UK MPs - this means we collected all the tweets sent by each current MP, any replies to those tweets, and any retweets either made by the MP or of the MP's own tweets. Note that this approach does not collect all tweets which an individual would see in their timeline, as it does not include those in which they are just mentioned. However, ``direct replies'' are included. We took this approach as the analysis results are more reliable  due to the fact that replies are directed at the politician who authored the tweet, and thus, any abusive language is more likely to be directed at them. Data were of a low enough volume not to be constrained by Twitter rate limits.

The study spans February 7th until May 25th 2020 inclusive, and discusses Twitter replies to currently serving MPs that have active Twitter accounts (574 MPs in total). Table~\ref{tab:corpus} gives the overall statistics for the corpus.

\begin{table}
\begin{tabular}{lrrrrrr}
Time period & Original & Retweet & Reply & ReplyTo & Abusive & \% Abuse\\
\hline
Total & 107,209 & 187,586 & 56,706 & 4,726,070 & 179,493 & 3.798\\
\end{tabular}
\caption{Corpus statistics. Columns give number of original tweets authored by MPs, number of retweets authored by them, number of replies written by them, number of replies received by them, number of abusive replies received by them, and abusive replies received as a percentage of all replies received.}
\label{tab:corpus}
\end{table}

\subsection{Rule-based Identification of Abusive Language}
\label{subsec:rules}
A rule-based approach was used to detect abusive language. An extensive vocabulary list of slurs (e.g. ``idiot''), offensive words such as the ``f'' word and potentially sensitive identity markers, such as ``lesbian'' or ``Muslim'', forms the basis of the approach. The slur list contained 1081 abusive terms or short phrases in British and American English, comprising mostly an extensive collection of insults, racist and homophobic slurs, as well as terms that denigrate a person's appearance or intelligence, gathered from sources that include \url{http://hatebase.org} and Farrell \textit{et al}~\cite{farrell2019exploring}. 131 offensive words were used, along with 451 sensitive words. ``Bleeped'' versions such as ``f**k'' are also included.

On top of these word lists, 53 rules are layered, specifying how they may be combined to form an abusive utterance as described above, and including further specifications such as how to mark quoted abuse, how to type abuse as sexist or racist, including more complex cases such as ``stupid Jew hater'' and what phrases to veto, for example ``polish a turd'' and ``witch hunt''. Making the approach more precise as to target (whether the abuse is aimed at the politician being replied to or some third party) was achieved by rules based on pronoun co-occurrence. 
Where people make a lot of derogatory comments about a third party in their replies to a politician however, for example racist remarks about others, there may be targeting errors leading to false positives.

The abuse detection method underestimates by possibly as much as a factor of two, finding more obvious verbal abuse, but missing linguistically subtler examples. This is useful for comparative findings, tracking abuse trends, and for approximation of actual abuse levels.

The method for detecting COVID-19-related tweets is based on a list of related terms. This means that tweets that are implicitly about the epidemic but use no explicit COVID terms, for example, ``@BorisJohnson you need to act now,'' are not flagged.

\subsection{Thematic Analysis}
\label{subsec:thematic_analysis}
To understand more about the kinds of tweets attracting high levels of abuse, we considered several approaches for ranking them. Ranking tweets by the most replies will surface prominent individuals, but perhaps not always polarising individuals or viewpoints. We decided on an initial criterion that a tweet had to have received \textbf{8\% or more abusive replies}, which is nearly twice the average level of abuse noted by~\cite{ward2017turds,gorrell2018twits,gorrell2020online}. In addition, to filter out tweets that were attracting just a handful of abusive replies, we examined tweets that received \textbf{at least 20 abusive replies}. 

All tweets were first examined and \textbf{coded openly}, as suggested by Braun and Clarke~\cite{braun2012thematic}, to see which patterns emerge. \textbf{Contentious subjects} (such as Brexit, racism, and even Jeremy Corbyn), as well as \textbf{potential media agendas} (such as reaching out to undervalued communities, or making criticisms of the party in government) emerged from this analysis. We then compared open codes to themes from the literature regarding \textbf{social media activities} that politicians may undertake during a health crisis, as well as \textbf{ideological themes and existing priors} that may be influencing how UK MPs are perceived. We created a final set of categories through the processes of reduction and comparison across the codes. We then re-coded the data according to the final annotation scheme in Table \ref{tab:mediacodes}.



\begin{table}
\begin{tabular}
{|p{1.5cm}|p{1cm}|p{3cm}|p{5cm}|}
\hline
\textbf{Media Category} & \textbf{No. Tweets} & \textbf{Description} & \textbf{Examples} \\
\hline
Defending & 14 & MPs responding to critiques of oneself or others & \allowbreak Replying to Keir Starmer: A low blow and misjudged at this time. Yes we need everyone to have PPE but the right use of It is also needed. - Stuart Anderson\\
\hline
Defending (e) & 3 & Similar to the above, but with ``escalation indicators'' (described below) & For God's sake. The man nearly died. He is going to Chequers away from the glare of publicity to recuperate and be with his partner who is due to give birth in a matter of weeks. Get a life - idiot keyboard warrior! \#HardHearted - Andrea Jenkyns\\
\hline
Direct Rebuke & 3 & MPs critiquing someone who is not directly an authority & No celebrations today. Only relief that the disastrous Corbyn era is over and hope that we can turn the page on what it did to Labour. Congratulations to (Kier Starmer) and (Angela Rayner) on their elections as Leader and Deputy Leader. - Pat McFadden\\
\hline
Direct Rebuke (e) & 3 & Similar to the above, but with escalation indicators & Not that I should be surprised by the lazy left but interesting how work-shy socialist and nationalist MPs tried to keep the remote Parliament going beyond 2 June. - Henry Smith \\
\hline
Direct Rebuke of Authorities & 60 & MPs critiquing people in power, in particular coming from opposition parties & It is irresponsible and short-sighted from the government to rule out extending the post-Brexit transition period. We should be taking action now to provide certainty for business in the face of this global economic challenge. - Lisa Nandy\\
\hline
Direct Rebuke of Authorities (e) & 70 & Similar to the above, but with escalation indicators & Boris Johnson boasting about shaking hands with coronavirus patients. You could not make it up. Britain is about to learn the hard way this is not the man to lead us in a crisis. - David Lammy\\
\hline
Engage Voters & 47 & MPs reaching out to potential new voters and speaking to core voters & If there's one clip to watch from \#Brits2020 last night, it's this. (Santan Dave) speaking truth to power. Hero - Zarah Sultana\\
\hline
Escalation & 9 & MPs making statements that just agitate & How many Johnson children are there now? - Barry Sheerman \\
\hline
Events & 12 & The MP's tweet relates back to an event that preceded the Tweet or a clear pattern of behaviour & NEWS: My Telegraph article on the next stage of our \#coronavirus plan: We must all do everything in our power to protect lives. - Matt Hancock \\
\hline
Information & 4 & MPs providing information without commentary & Data sources and maps here: - Richard Burgon \\
\hline
Proactive & 24 & MPs relating a sense of doing something about the problem  & Dear @patel4witham please do the right thing and release the women in Yarlswood now. - Jess Phillips\\
\hline
Unclear & 4 & Tweet could not be annotated or belongs to no clear category & Quite the change in rhetoric from the days when Welsh Government were encouraging people to come to Wales and drive their 4x4's right onto the beaches. - Stephen Crabb\\
\hline
\textbf{Grand Total} & 190 &  &  \\
\hline

\end{tabular}

\caption{Media Categories that express MPs' activities on Twitter, to which the MP received abusive replies. Descriptions and examples included.}
\label{tab:mediacodes}
\end{table}

In addition to these codes, we also examined the topics MPs referred to in their posts and grouped them inductively into categories of topics that are considered controversial in UK politics. These were: \textbf{home rule/nationalism with respect to Northern Ireland, Scotland or Wales; inequality; and Brexit, alongside specific individuals and the ways in which they communicate}. Of course, the topic of \textbf{COVID-19 itself - the government response and the impacts-} was a primary topic category as well. Following the distribution of these codes and topics in our sample, we used descriptive statistics and further qualitative analysis to expand on trends and observations uncovered through our computational approaches. 

In consideration of rigour, we have taken several steps to adjust for having used one annotator for the qualitative analysis. Barbour has suggested focusing on alternative explanations in analysis, rather than a potentially superficial measure of inter-rater agreement through multiple coders~\cite{barbour2001checklists}. In addition, we provide a full justification of the coding scheme against each tweet in the sample, available at the URL noted in Section \ref{sec:declarations}, so that other researchers can interrogate and interpret our findings accordingly. \\

In the following section, we present our findings, which are organised by research question and which include both the quantitative and qualitative data analysis that answer that particular research question. 

For RQ1, we rely primarily on our literature review of trends and high-level analysis of abuse toward British MPs. We compare this with our findings from the COVID-19 period we studied. For RQ2, we consider how our findings fit the dimensions noted by Andreouli \textit{et al}~\cite{andreouli2019brexit} as impacting contemporary British political discourse: \textbf{ideology, authority and affect}. We explore the four factors that contribute to how these dimensions are perceived, such as prominence, specific events, engagement habits and features of identity~\cite{gorrell2020politicians}. Finally, for RQ3, we present our qualitative analysis on the social media activities of UK MPs during COVID-19 thus far. We contextualise the abusive responses they receive for these activities, given the dimensions and contributing factors that may play a role. 

\section{Findings: General Trends and Comparisons (RQ1)}
\label{findings:generaltrends}

Our first research question asked: \textit{How has the context of COVID-19 impacted the typical patterns that have been observed in previous work about hateful and abusive language toward UK MPs?} To answer this question, we begin with a review of the time period studied, namely February 7th until May 25th 2020 inclusive, placing it in historical context.


Gorrell et al~\cite{gorrell2020mp} use the same (cautious) abuse counting methodology as we use here to show that aside from a blip around the 2015 general election, abuse toward MPs on Twitter has been tending to rise from a minimum of 2\% of replies in 2015, peaking mid-2019 at over 5\% with a smaller peak of around 4.5\% around the 2019 general election. After the election, however, abuse toward MPs fell to around 3.5\%.

\begin{figure}
  \includegraphics[width=.85\textwidth]{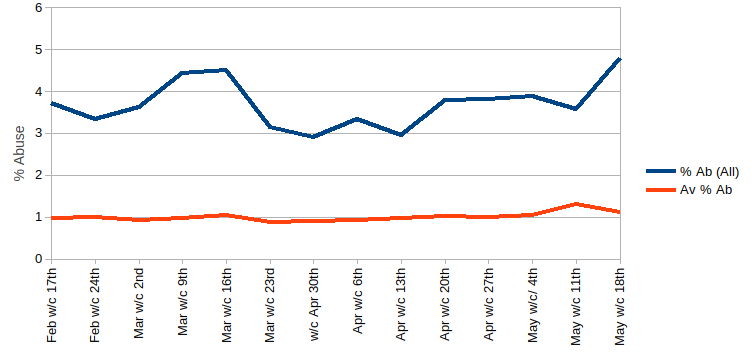}
  \caption{Abuse percentage received by all MPs, macro- (red) and micro- (blue) average, per week}
  \label{fig:timeline-per-week}
\end{figure}

In the timeline in Fig~\ref{fig:timeline-per-week} we zoom in on the study period, and show abuse levels overall, toward all MPs, on a per-week basis since mid-February. This timeline shows a rise in abuse, back up to over 4\% around the time of the introduction of social distancing, before dipping, and then gradually beginning to rise again later in the study period. We see that the macro-average abuse level (red line) remains relatively steady, suggesting that this fluctuation is confined to a small number of high profile politicians (therefore being more evident in the micro-averaged blue line).

\begin{figure}
  \includegraphics[width=.95\textwidth]{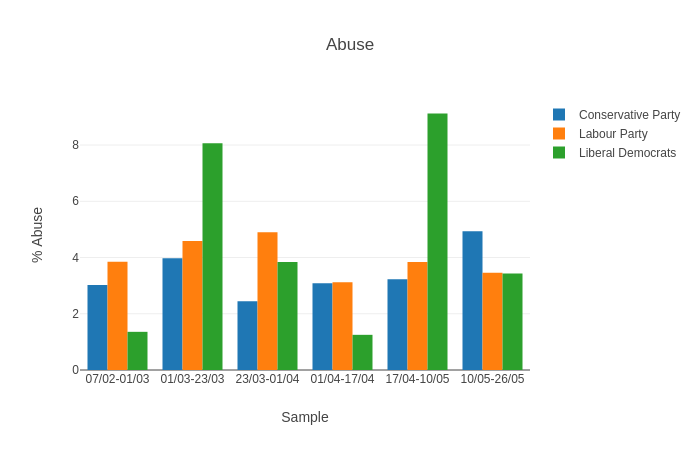}
  \caption{Abusive replies as a percentage of all replies received, micro-average, split by party and time period.}
  \label{fig:parties}
\end{figure}

Fig~\ref{fig:parties} shows abuse received as a percentage of all replies received by MPs, for six distinct time periods discussed in more detail below. We see that on the whole, response to the Conservative party has been favourable. The exception is after May 10th, when the negative response to Dominic Cummings' decision to travel north with COVID-19 symptoms came to the fore. Responses to Liberal Democrat MPs are more erratic due to their lower number. In previous studies, we have found Conservatives receiving higher abuse levels, yet here we see Labour politicians receiving more abuse in most periods. This was in evidence even in February, so precedes the pandemic, although Twitter has tended to be left-leaning in the UK~\cite{gorrell2019local}. It remains to be seen if this is the beginning of a swing to the right or if it is specific to the times, e.g. arising from a desire to trust authority during times of crisis \cite{volkan2014blind}.

There is a significant negative correlation between receiving a high level of COVID-related attention and receiving abuse (-0.52, p\textless0.001, Feb 7th to May 25th, Spearman's PMCC). We see this clearly in prominent government figures below, who are receiving the lion's share of the COVID-19 attention and lower levels of abuse than seen for them in pre-COVID periods~\cite{gorrell2018twits,gorrell2020politicians}. However the correlation is significant across the sample of all MPs. The reaction of the public to the Conservative party and the government's actions during COVID-19 may be related to the conditions of a public health crisis as discussed in~\cite{veil2011work,llewellyn2020COVID}, in which citizens may feel more motivated to trust authorities, although it may also follow from the crisis engaging a different group of people than usually respond to politicians on Twitter.

With a view to separating out different groups of Twitter users, we tracked hashtags relating to dominant pro- vs anti-lockdown perspectives, as well as issues of concern; namely conspiracies and misinformation, and racism in conjunction with the pandemic. Pro- and anti-lockdown hashtags were easily acquired, being dominant hashtags appearing in the dataset. They were then extended with minor linguistic variants. This report\footnote{\small{\url{http://moonshotcve.com/COVID-19-conspiracy-theories-hate-speech-twitter/}}} from Moonshot CVE was used as a guide to the overall conspiracy landscape within COVID-19. They provide some hashtags, and variants were then acquired, again, from looking down the list of hashtags appearing in the dataset for other variants, and including linguistic variations. The areas they highlight are anti-Chinese feeling/conspiracy theory, theories that link the virus to a Jewish plot, theories that link the virus to an American plot, generic ``deep state'' and 5g-based theories and general theories that the virus is a plot or hoax. Table~\ref{tab:hashtags} shows substantial evidence of ill-feeling toward China.

In our analysis, MPs using Chinese data or referencing the Chinese government's response to COVID-19 in a positive context appear to attract abuse. One example, in terms of receiving a high percentage of abuse as well as a notable degree of attention, was the one below from Richard Burgon.

\url{https://twitter.com/RichardBurgon/status/1244043022297370626} (17\% abuse, or 6\% of all abuse sent to MPs in March post-lockdown):

\begin{quote}

\textit{This is a Trump-style attempt to divert blame from the UK government's failures.}

\textit{A World Health Organization report says China ``rolled out perhaps the most ambitious, agile \& aggressive disease containment effort in history''}

\textit{We haven't even sorted out enough tests for NHS staff}
\end{quote}

China's record of human rights or transparency is often provided as evidence of argument against such tweets. However, mixed in with this are also a number of Sinophobic comments about China having ``caused'' or ``started'' the virus, for which at present there is no reliable scientific evidence. It is not clear how potentially useful critiques of the Chinese government may be discussed without also provoking more sinister, racist commentary.

Classic conspiracy theories are in evidence but numbers of mentions are low (though note that most of the 183 mentions of ``NWO'' (``new world order'') are now COVID-19-related, suggesting opportunistic incorporation of COVID-19 into existing mythologies). There is considerable evidence of some Twitter users not believing in the virus, and numbers of mentions to this effect are within one order of magnitude of the popular ``stay home save lives''. Yet all are surpassed by the theme of economic support for those not in established employment ("\#newstarterfurlough").

\begin{table}
\resizebox{\textwidth}{!}{%
\begin{tabular}{lrrr}
\textbf{Search terms (in all replies to MPs, not case-sensitive)} & \textbf{\# tweets} & \textbf{\# abusive} & \textbf{\% abusive}\\
\hline
``\#endthelockdown''  & 5,506 & 272 & 4.940\\
``\#newstarterfurlough'' OR ``\#newstarterjustice'' & 55,593 & 243 & 0.437\\
``\#stayhomesavelives'' & 20,538 & 1222 & 5.950\\
``\#coronahoax'' OR ``\#hoaxvirus'' OR ``\#fakevirus''  & 415 & 61 & 14.699\\
``\#coronabollocks''  & 250 & 19 & 7.600\\
``\#plandemic'' OR ``\#scamdemic'' OR ``\#fakepandemic'' & 1,178 & 64 & 5.433\\
``\#filmyourhospital'' OR ``\#emptyhospitals'' & 59 & 5 & 8.475\\
``\#ccpvirus'' OR ``\#chinaliedpeopledied'' & 2,463 & 38 & 1.543\\
``\#chinesevirus'' & 2,208 & 119 & 5.389\\
``\#NukeChina'' OR ``\#DeathtoChina'' OR ``\#DestroyChina'' & 31 & 2 & 6.452\\
``\#GatesVirus'' OR ``\#CIAVirus'' OR ``\#AmericaVirus'' & 69 & 2 & 2.899\\
``\#5gcoronavirus'' & 53 & 1 & 1.877\\
\hline
\end{tabular}%
}
\caption{Mention count of viewpoint-related hashtags, in all replies to MPs, Feb 7th to May 25th inclusive. Some further variants of the terms given, including non-hashtag mentions in text, are also included but not listed here for brevity; see Gorrell et al~\cite{gorrell2020mp} for a more complete description}
\label{tab:hashtags}
\end{table}

So across the board, \textbf{COVID-19 has not led to higher proportions of abuse on Twitter for MPs compared with the high levels of abuse directed at them in 2019}. However, these findings might be partially explained by \textbf{varying degrees of engagement} by different societal groups, in addition to \textbf{events affecting attitudes} to authority. As we will see in Section~\ref{sec:politicalauthority}, the comparatively positive response to Boris Johnson might be explained by more people replying to him than would normally do so; this extra attention was not abusive. The lower levels of abuse received by MPs who receive more tweets mentioning COVID-19 might also be explained by \textbf{different people replying to politicians than usually would}. A particularly striking illustration of this comes from tweets to MPs using the hashtag \#newstarterfurlough and variants. People who had recently started a new job "fell through the cracks" for financial support from the government. With 56,000 tweets to MPs using \#newstarterfurlough and variants (compared with only 21,000 using \#stayhomesavelives and variants), \#newstarterfurlough is the dominant hashtag campaign of the period. Given that those individuals are in an unfortunate position, it is all the more surprising to find that only 0.4\% of those tweets contained abuse, as shown in Table~\ref{tab:hashtags}. A possible explanation is that the ``new starters" are a \textbf{broader, and more polite, cross-section of society} than people who usually reply to politicians on Twitter.

In contrast, tweets containing \#stayhomesavelives and variants contained 6.0\% abuse. Tweets containing hashtags refuting the very existence of the virus, for example \#scamdemic and \#hoaxvirus, contained 7.8\% abuse. Tweets describing COVID-19 as "Chinese", e.g. containing \#chinesevirus, contained 5.4\%. Tweets found containing anti-lockdown hashtags contained 4.9\% abuse.

\section{Dimensions of Political Discourse During COVID-19 (RQ2)}
\label{findings:dimensions}
Our second research question asked \textit{what are the societal dimensions that appear to impact how media activities are perceived during the COVID-19 pandemic and how does this compare with those that impacted Brexit or recent general elections?}. For this question, we drew on previous work described in section \ref{sec:RelatedWork}, more specifically in \ref{subsec:HateBritishMPs} on levels of abuse toward UK MPs. We then examined the time period covered by our study in the context of three dimensions drawn from Andreouli \textit{et al} \cite{andreouli2019brexit}: \textbf{political authority, ideology and affect}. We use the 4 factors from Gorrell \textit{et al} \cite{gorrell2020politicians} to help further describe these dimensions in terms of: \textbf{prominence, event surges, engagement and identity}. 

\subsection{Political Authority}
\label{sec:politicalauthority}

As mentioned previously, Brexit created a notion of political authority that presented sovereignty of the UK on one side or community with Europe on the other~\cite{andreouli2019brexit}. During the past three UK elections, partisanship has led to a splintering of political authority and an erosion of trust~\cite{gorrell2019partisanship}. During the COVID-19 pandemic, we can see a similar effect, for example, the wearing of face-coverings as a personal \textit{versus} social choice\footnote{https://www.southwalesargus.co.uk/news/18444799.wearing-face-covering-mask-wales-not-compulsory/}, and participation in protest \textit{versus} public health.\footnote{https://www.forbes.com/sites/tarahaelle/2020/06/19/risking-their-lives-to-save-their-lives-why-public-health-experts-support-black-lives-matter-protests/\#11b5ac96851b}

Increased name recognition and popularity have also been associated with higher levels of abuse in both the cases of Brexit and General Elections~\cite{mcloughlin2017turds,ward2020turds,gorrell2018twits,gorrell2020politicians}. Our COVID-19 data shows similar trends.

\begin{figure}
  \includegraphics[width=.85\textwidth]{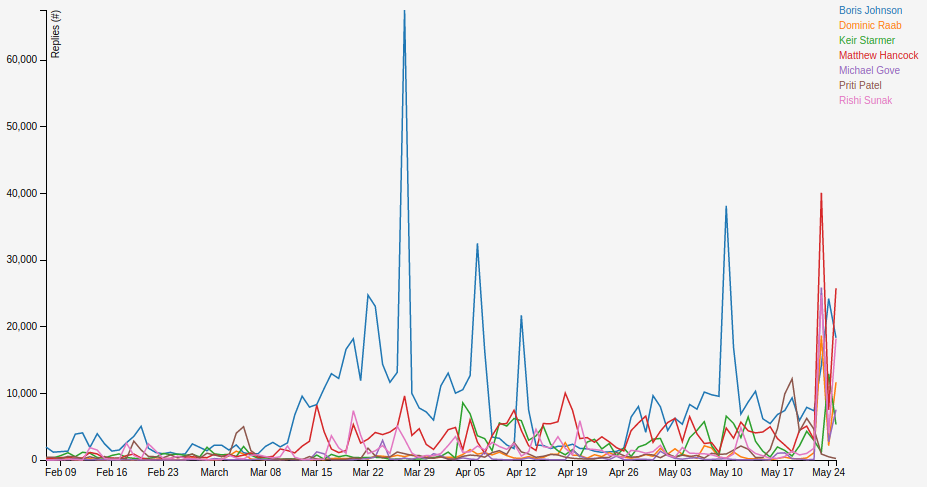}
  \caption{Number of replies received by relevant ministers and opposition leader Keir Starmer per day from February 7th to May 25th 2020 inclusive.}
  \label{fig:timeline}
\end{figure}

Fig~\ref{fig:timeline} shows number of replies to prominent politicians since early February, and shows that for the most part, \textbf{attention during COVID-19 has focused on Boris Johnson}. He received a large peak in Twitter attention on March 27th. 58,286 replies were received in response to his tweet announcing that he had COVID-19. Abuse was found in 2.3\% of these replies. This is \textbf{low for a prominent minister} as we may discern from Fig~\ref{fig:timeline-per-week}, indicating a generally supportive response to the prime minister's illness. Further peaks on Mr Johnson's timeline correspond to the dates on which he was admitted to intensive care (April 6th), left hospital to recuperate at Chequers (April 12th), and began to ease the lockdown (May 10th). The late burst of attention on other politicians arises from several tweets by ministers in support of Dominic Cummings, the senior government advisor who chose to travel north to his parents' home in the early stages of his illness with COVID-19.

However, one tweet receiving a high level of abuse regarded the very first video address made by Boris Johnson in response to the pandemic:

\url{https://twitter.com/BorisJohnson/status/1238365263764041728} (9\% of replies were abusive, tweet received 3\% of all abuse to MPs in the period. It also includes a video statement.)

\begin{quote}
\textit{This country will get through this epidemic, just as it has got through many tougher experiences before.}
\end{quote}

For those who trust in Boris Johnson's leadership and who like the way he communicates, this tweet may have provided some comfort. A review of replies to this tweet shows that supporters tweeted messages of appreciation and hope in response. For those who do not trust him and who believe he should have acted sooner, this tweet was perceived as a provocation. Several replies that are critical but not abusive point to official sources of information from elsewhere in Europe, or make advisement to the public about staying home and avoiding non-essential journeys. 

In this sense, \textbf{COVID-19 lends its own political authority} to some arguments. However, invoking COVID-19 as a member of the opposition, especially in the context of persistent debates, is often met with accusations of \textbf{``playing party politics''}. The following tweet by David Lammy received 16\% abusive replies, representing 1\% of all abuse sent to MPs in the March pre-lockdown period:

\url{https://twitter.com/DavidLammy/status/1239835712444391424} 

\begin{quote}
\textit{No more government time, energy or resources should be wasted on Brexit this year. Boris Johnson must ask for an extension to the transition period immediately. \#COVID19 is a global emergency.}
\end{quote}

Several MPs made comments about the need to extend the Brexit transition period. All received a considerable amount of replies containing abusive language.  

More generally, there is disagreement about the \textbf{priorities of government during a crisis}. For example, there was a considerable amount of abuse directed at Richard Burgon for a tweet in which he discussed the accomplishments of his work as shadow justice secretary.\footnote{https://twitter.com/RichardBurgon/status/1247248198932062208} This was regarded by some critics as mistimed,  given the PM's health at the time, and received 17\% abuse, constituting 2\% of all abuse sent to MPs between April 1st and 16th inclusive.

In later sections, we will explore how the \textbf{media activities of opposition parties} are perceived by the public, in particular in connection with contentious subjects.

\subsection{Events}
\label{sec:events}

Events are somewhat in a category of their own, as \textbf{an MP's past actions and words} become part of the \textbf{public's priors} in understanding the position of that MP. To contextualise the level of attention to MPs and the abusive replies they received in connection with specific events, 
we review the events of the period, both in terms of who is receiving abusive replies and in the themes that are rising during the period as demonstrated by the appearance of certain hashtags. 

\subsubsection{Beginning of the Pandemic}

\begin{figure}
  \includegraphics[width=.75\textwidth]{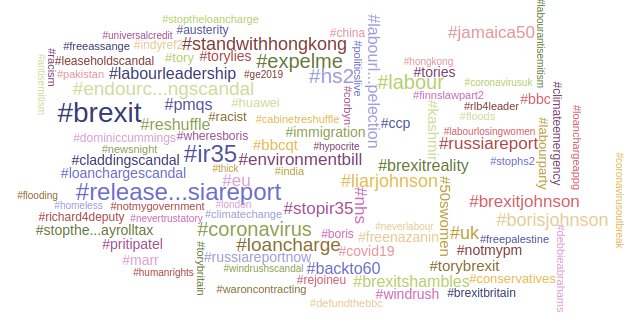}
  \caption{Top 100 hashtags in all replies sent to MPs - February 7-29 2020 inclusive.}
  \label{fig:feb-7-cloud}
\end{figure}

The hashtag cloud in Fig~\ref{fig:feb-7-cloud} 
shows that Brexit remained the dominant topic in Twitter political discourse during February, with the epidemic not yet having arrived in the UK.

Table~\ref{tab:feb-7} gives a baseline for attention on MPs as we go into the pandemic, showing that aside from Boris Johnson, attention, and abuse, is high for Labour politicians. The column “Authored” refers to the number of tweets originally posted from that account that were not retweets or replies. ``replyTo'', refers to all of the replies received to the individual’s Twitter account in that period. The next column, ``COV'', is the number of replies received to that account containing an explicit mention of COVID-19, with the following column representing the number of replies that verbal abuse was found in (``Abusive''). The last three columns present the data in a comparative fashion. Firstly, we have the percentage of replies that the individual received that were abusive. Next, we have the percentage of replies that were COVID-related. The last column is the percentage of COVID-related replies to that individual, in comparison with all COVID-related replies received by all MPs.

\begin{table}
\resizebox{\textwidth}{!}{%
\begin{tabular}{lrrrrrrr}
\textbf{Name} & \textbf{Authored} & \textbf{replyTo} & \textbf{COV} & \textbf{Abusive} & \textbf{\% Ab} & \textbf{\% COV} & \textbf{\% Total COV}\\
\hline
\cellcolor{blue!15}Boris Johnson & 14 & 48,379 & 1,072 & 1,695 & 3.504 & 2.216 & 37.773\\
\cellcolor{red!15}David Lammy & 89 & 47,368 & 73 & 2,308 & 4.872 & 0.154 & 2.572\\
\cellcolor{red!15}Richard Burgon & 184 & 30,789 & 18 & 1,556 & 5.054 & 0.058 & 0.634\\
\cellcolor{red!15}Jeremy Corbyn & 25 & 29,550 & 215 & 2,400 & 8.122 & 0.728 & 7.576\\
\cellcolor{red!15}Rebecca Long-Bailey & 121 & 27,113 & 17 & 823 & 3.035 & 0.063 & 0.599\\
\cellcolor{red!15}Zarah Sultana & 96 & 18,630 & 12 & 677 & 3.634 & 0.064 & 0.423\\
\cellcolor{red!15}Debbie Abrahams & 96 & 18,186 & 45 & 822 & 4.520 & 0.247 & 1.586\\
\cellcolor{red!15}Keir Starmer & 97 & 15,455 & 26 & 360 & 2.329 & 0.168 & 0.916\\
\cellcolor{red!15}Lisa Nandy & 110 & 15,271 & 9 & 413 & 2.704 & 0.059 & 0.317\\
\cellcolor{blue!15}Priti Patel & 9 & 13,664 & 45 & 345 & 2.525 & 0.329 & 1.586\\
\end{tabular}
}
\caption{MPs with greatest number of replies from February 7 - 29 2020 inclusive. Cell colours indicate party membership; blue for Conservative, red for Labour.}
\label{tab:feb-7}
\end{table}

\begin{figure}
  \includegraphics[width=.85\textwidth]{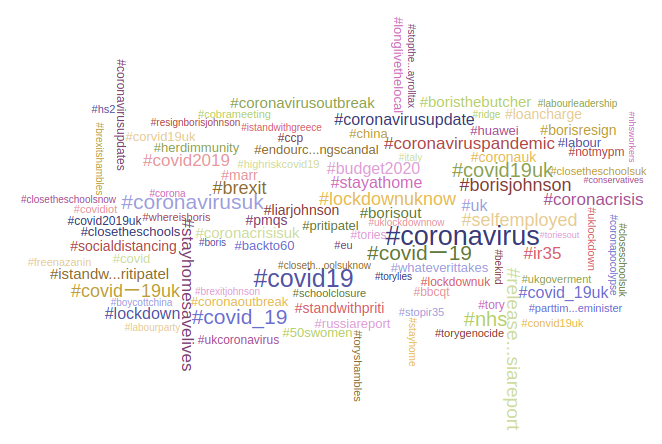}
  \caption{Top 100 hashtags in all replies sent to MPs - March 1st - 22nd 2020 inclusive.}
  \label{fig:mar-1-cloud}
\end{figure}

The word cloud in Fig~\ref{fig:mar-1-cloud} shows all hashtags in tweets to MPs in earlier part of March, and unsurprisingly shows a complete topic shift, to the subject of the epidemic, to the virtual exclusion of all else.

\begin{table}
\resizebox{\textwidth}{!}{%
\begin{tabular}{lrrrrrrr}
\textbf{Name} & \textbf{Authored} & \textbf{replyTo} & \textbf{COV} & \textbf{Abusive} & \textbf{\% Ab} & \textbf{\% COV} & \textbf{\% Total COV}\\
\hline
\cellcolor{blue!15}Boris Johnson & 45 & 160,356 & 28,818 & 7,684 & 4.792 & 17.971 & 49.380\\
\cellcolor{red!15}David Lammy & 120 & 43,386 & 1,602 & 2,952 & 6.804 & 3.692 & 2.745\\
\cellcolor{blue!15}Matthew Hancock & 100 & 42,520 & 7,167 & 1,800 & 4.233 & 16.856 & 12.281\\
\cellcolor{red!15}Jeremy Corbyn & 44 & 42,435 & 1,824 & 3,244 & 7.645 & 4.298 & 3.125\\
\cellcolor{blue!15}Rishi Sunak & 47 & 25,534 & 2,225 & 284 & 1.112 & 8.714 & 3.813\\
\cellcolor{blue!15}Nadine Dorries & 52 & 24,731 & 1,275 & 466 & 1.884 & 5.155 & 2.185\\
\cellcolor{red!15}Jess Phillips & 154 & 20,931 & 394 & 541 & 2.585 & 1.882 & 0.675\\
\cellcolor{red!15}Lisa Nandy & 74 & 20,355 & 234 & 925 & 4.544 & 1.150 & 0.401\\
\cellcolor{red!15}Richard Burgon & 145 & 20,281 & 446 & 1,222 & 6.025 & 2.199 & 0.764\\
\cellcolor{red!15}Zarah Sultana & 107 & 18,796 & 204 & 815 & 4.336 & 1.085 & 0.350\\
\end{tabular}
}
\caption{MPs with greatest number of replies from from March 1st - 22nd 2020 inclusive. Cell colours indicate party membership; blue for Conservative, red for Labour, yellow for Liberal Democrat.}
\label{tab:mar-1}
\end{table}

We see from Table~\ref{tab:mar-1} that with the arrival of COVID-19 in the UK, Health secretary Matt Hancock became more prominent on Twitter at this time, though attention was not more abusive. Attention on chancellor Rishi Sunak also increased and was not abusive. This is consistent with previous research that indicates a public willingness and desire to trust authorities in a crisis~\cite{veil2011work,llewellyn2020COVID}. We see a high level of attention on Boris Johnson, but the abuse level is lower than was seen for him in previous years (we found 8.39\% in the first half of 2019 as mentioned above; in 2017 as foreign secretary Mr Johnson received similarly high abuse levels in high volumes). Negative attention on Labour politicians is high, but note that this was also the case before the start of the epidemic in the UK.





Matt Hancock received 11\% abusive replies (3\% of all abuse to MPs in the March pre-lockdown period) to the following tweet, in which he released his Telegraph article on the government's response to the virus.

\url{https://twitter.com/MattHancock/status/1238960146342084609} 

\begin{quote}
\textit{NEWS: My Telegraph article on the next stage of our \#coronavirus plan: \textless link\textgreater}

\textit{We must all do everything in our power to protect lives}
\end{quote}

Critics were angry that Mr. Hancock would post important government information, during a time of extreme uncertainty, behind a pay wall. This goes back to the public's information-seeking needs during a health crisis~\cite{veil2011work,llewellyn2020COVID}. Not only is information important for collective sense-making, it is also important for determining personal risk~\cite{vos2016social}.

\subsubsection{Beginning of Lockdown}

With the commencement of lockdown, the rise of the hashtags ``\#stayhomesavelives'' and \#lockdownuknow shows a shift toward comment on the practical details.
Support for the lockdown appears to be high at this stage, with the top ten hashtags featuring only pro-lockdown or generic COVID-19 tags. Attention continues to focus on Boris Johnson (see Gorrell et al~\cite{gorrell2020mp} for complete word clouds and tables as well as histograms for each period), and is even less abusive than previously, largely due to a surge in non-abusive attention in conjunction with his being diagnosed with COVID-19. By volume, the most abuse-generating tweet was Boris Johnson's illness announcement, but as a percentage this was remarkably un-abusive, as discussed above, with only 2.3\% abuse. The high abuse count follows only from the very high level of attention this tweet drew.



Moving into April 2020, the rise of the hashtag ``\#newstarterfurlough'' shows that prior to Donald Trump's ``liberation'' tweets and the visible emergence of an anti-lockdown backlash, attention has already begun to focus on the economic cost of the lockdown, as illustrated by the prominence of hashtags such as \#newstarterfurlough and \#wearethetaxpayers. 


Boris Johnson's abuse level continues to be low as his illness takes a serious turn. In the context of the pandemic, different influences from the public also have a measure of authority. For example, the high abuse level toward Jack Lopresti during this period relates to his controversial opinion that churches should open for Easter. Strong opinion in conjunction with a religious event is part of a pattern that we see in conjunction with Eid in the next section.

\url{https://twitter.com/JackLopresti/status/1247508135029411841} (18\% abuse, 6\% of all abuse sent to MPs between April 1st and 16th inclusive):

\begin{quote}
\textit{Open the churches for Easter – and give people hope \url{https://telegraph.co.uk/news/2020/04/06/open-churches-easter-give-people-hope/?WT.mc_id=tmg_share_tw…} via @telegraphnews}
\end{quote}

\url{https://twitter.com/JackLopresti/status/1247894726486798342} (17\% abuse, 3\% of all abuse sent to MPs between April 1st and 16th inclusive):

\begin{quote}
\textit{Today I wrote to The Secretary of State @mhclg and also sent a copy of this letter to Secretary of State @DCMS to ask the Government to consider opening church doors on Easter Sunday for private prayer.}
\end{quote}




\subsubsection{Lockdown Backlash}

From mid-April 2020, a notable backlash against lockdown began to emerge. Hashtags now appear to be critical, often economically focused but also including accusations of lying against China, Boris Johnson and Conservatives, and references to the shortage of personal protective equipment for medical workers. The distinct change in tone echoes events in the USA.\footnote{\small{E.g. \url{https://www.theguardian.com/global/video/2020/apr/16/armed-protesters-demand-an-end-to-michigans-coronavirus-lockdown-orders-video}}}


In this context it is interesting therefore that the tweet receiving the most abusive response by volume (it also received a striking level of abuse by percentage) is this one by Ed Davey.

\url{https://twitter.com/EdwardJDavey/status/1253882262715842560} (19\% abuse, 5\% of all abuse toward MPs for the period):

\begin{quote}
\textit{A pre-dawn meal today}

\textit{Preparing for my first ever fast in the holy month of Ramadan}

\textit{For Muslims doing Ramadan in isolation, you are not alone!}

\textit{\#RamadanMubarak}

\textit{\#LibDemIftar}
\end{quote}

The following tweet also attracted high levels of abuse by volume:

\url{https://twitter.com/jeremycorbyn/status/1253341601599852544} :

\begin{quote}
\textit{Ramadan Mubarak to all Muslims in Islington North, all across the UK and all over the world.}
\end{quote}

This tweet received 11\% abusive replies, 2\% of abuse for the period - this was also St George's Day, so perceived as evidence of anti-English sentiment, as in the following paraphrased replies for example: ``@jeremycorbyn So nothing about St George's day then? Ah, that's because we are English, the country you wanted to run but hate with a vengeance. And you wonder why you suffered such a huge defeat at the election'' and ``@jeremycorbyn So no mention of St. George's day then? You utter cretin.''

These attempts to reach voters and how they are perceived by those not within the same ideological framework will be discussed in Section \ref{findings:dimensions}.

\subsubsection{Lifting of Restrictions}
\label{subsub:May10}

As lockdown begins to be eased in May 2020, we see a return to a high level of focus on Boris Johnson, with 194,000 replies compared to Matt Hancock's 124,000 as the next most replied-to MP. Other senior Conservatives are also prominent. High levels of abuse are received by ministers who defended Dominic Cummings' actions on Twitter; Matthew Hancock (6.37\%), Oliver Dowden (8.17\%) and Michael Gove (7.13\%). Boris Johnson also receives more abuse than he did in the previous period (4.92\%). Example tweets are given below of ministers defending Mr Cummings:\footnote{More tweets by ministers defending Mr Cummings:\\ https://twitter.com/OliverDowden/status/1264221876374646786\\https://twitter.com/michaelgove/status/1264126108733186050}

\url{https://twitter.com/MattHancock/status/1264162359733555202} (7\% abuse, 7\% of abuse for the period):

\begin{quote}
\textit{I know how ill coronavirus makes you. It was entirely right for Dom Cummings to find childcare for his toddler, when both he and his wife were getting ill.}
\end{quote}





\url{https://twitter.com/MattHancock/status/1264975804208947208} (9\% abuse, 4\% of abuse for the period):

\begin{quote}
\textit{Dom Cummings was right today to set out in full detail how he made his decisions in very difficult circumstances. Now we must move on, fight this dreadful disease and get our country back on her feet}
\end{quote}

Hashtags show a high degree of negative attention focused on the partial treatment of Dominic Cummings, whilst continued attention on the economic plight of new starters is also in evidence. This signals the beginnings of contention that blossom in the periods after this study was concluded.  


\subsection{Ideology}
The ideologies that influenced Brexit, such as anti-prejudice and tolerance~\cite{andreouli2019brexit} are still apparent in the context of COVID-19. \textbf{``Virtue signalling''} is a common complaint attached to nearly every tweet that addresses issues of inequality and some that are trying to engage voters (see Section \ref{findings:socialmediaactivities}). Virtue signalling is defined as behaviour that indicates support for causes or sentiments that carry moral value, such as donating to charity~\cite{wallace2020consuming}, without much actual effort or care for the topic behind it. Disagreement on what constitutes virtue signalling \textit{versus} actually caring about a social issue creates a ``liminal hotspot'' where misunderstanding takes place~\cite{andreouli2019brexit}. We go more deeply into the subject of virtue signalling in Section \ref{findings:dimensions}.

In many cases, \textbf{identity and ideology are interrelated through experience}. Individuals from minority backgrounds speak about racism more often, also in context of COVID-19, potentially because they experience it. When MPs from under-valued or under-represented minorities speak to their voters about racism, they are not only speaking to voters who need to understand experiences of racism, but also to voters who experience it directly. If the MP has a track record in working toward racial justice, their election is a signal that they should keep doing this work. Of the 15 tweets shared by Women of Colour in our qualitative sample (190 Tweets receiving a high percentage and number of abusive replies), more than 50\% are about engaging voters and being proactive toward issues of inequality (we see this in more detail in \ref{subsec:contentiousissues} and in Table~\ref{fig:touchy-subjects}). Another 40\% are direct rebukes of authority from women in opposition parties. It seems that Women of Colour are disproportionately carrying the flame for the highly abused topic of inequality, as we see in Fig~\ref{fig:touchy-subjects}. This may have partly to do with the party they belong to, but it appears to be also partly about the topics and expertise these women bring to the table. For example, Bell Ribeiro-Addy was elected for the first time in December 2019 after a long career of addressing inequality in migration. She addresses Sinophobia in communications about coronavirus and its origins in this Tweet:

\url{https://twitter.com/BellRibeiroAddy/status/1247570145733758980}
\begin{quote}
\textit{As senior Conservatives publish Sinophobic screeds in the rw press to distract from their own Government’s lethal complacency, it’s clear racism won’t stop for \#Coronavirus. Neither must we opposing it. Join the fightback in an hour! http://ow.ly/4GBm30qw1jX}   
\end{quote}

This quote communicates proactiveness, for example, with the words "oppose" or "fightback". As such, this is the kind of statement that is not meant for those who don't accept racism as a fact of experience, or who do not think it is an important issue in the context of COVID-19. It is a call to action for those who do.

This tweet, from MP Rupa Huq, addresses the lack of women ministers present at press briefings:   

\url{https://twitter.com/RupaHuq/status/1246107696652320769}
\begin{quote}
   \textit{Once again headed up by a man for the umpteenth time - who we only heard a week ago had tested \#COVID19 positive to boot. When will Downing Street allow a woman minister to front up one of these press shindigs?}
\end{quote}

Several responses to this tweet ask why the MP chooses to focus on gender, given that the roles most relevant during the COVID-19 pandemic just happen to be held by men. Or, they ask her to explain why she believes a woman could do a better job. Some responses also refer to this as ``virtue signalling", though the MP is a woman and her comment is about representation of women. 

In terms of the General Elections in 2015 and 2017, topics of concern were primarily around the economy, Europe and the NHS. While Europe and Brexit are still present as important subjects, our topic analysis and the hashtags collected in each period show that the economy is now the greatest concern for users on Twitter, along with various implications for public health, survival of businesses, unemployment, and social welfare. The significant financial support the government has provided during COVID-19 has revived discussions about socialism and capitalism more generally as economic models:

\url{https://twitter.com/jeremycorbyn/status/1238897340309790721}
\begin{quote}
\textit{There's no statutory sick pay for part-time, low-paid or zero-hours contract workers. And the rate of sick pay isn't enough to live on. Wrong at any time - but dangerous while people who might be ill are asked to stay home. The system is broken and now is the time to fix it.}
\end{quote}

This quote from Jeremy Corbyn attracted supportive messages, including some that are abusive toward Boris Johnson or other members of the Conservative party. However, there were also many criticisms of the tweet, which primarily express exasperation with Mr. Corbyn or chastise him for bringing this subject up in the middle of a crisis. If one agrees with Mr. Corbyn and feels that this crisis has only further exemplified the failings of the social welfare system in the United Kingdom, his words will resonate. 

\subsection{Affect}
Previous work has suggested that impassioned speech provides a measure of credibility on its own. MPs have been asked to consider the tone of their messages to the public and to one another in Parliament~\cite{distaso2015managing}. However, in COVID-19, in which the country is embroiled in a large public health crisis, it is difficult to determine exactly how tone impacts the political discourse. 
When we were looking for escalations in our qualitative sample, we looked at critiques of MPs' tweets to try and understand what someone might object to in a given statement. Typically, hyperbolic language, sarcasm, insult, and making something personal are the escalations that are named. However, to avoid potentially classifying a tweet as escalating for using strong language around events that are urgent matters of public health, we required at least two measures of escalation to be present in order to classify a tweet as a potential escalation. When examining our sample, escalations were present in about a quarter of the tweets. 

As a percentage of replies, a notable tweet was the following:

\url{https://twitter.com/HenrySmithUK/status/1263394101002674176} (13\% abuse, 2\% of abuse for the period of May):

\begin{quote}
\textit{Not that I should be surprised by the lazy left but interesting how work-shy socialist and nationalist MPs tried to keep the remote Parliament going beyond 2 June.}
\end{quote}

In the context of an increasingly uncertain economic situation for many individuals, respondents felt that Mr. Smith was accusing those who have been furloughed or who are shielding of avoiding work. Several respondents also implied that work in communities to provide social support was not being valued. Though Mr. Smith's comments (and his affect) may have been directed at his colleagues in Parliament, he hit a mark with left-leaning members of the British public as well. 

Many other tweets (n=60) are about rebuking authorities, which is sometimes done using strong language. We have 14 tweets from David Lammy in the sample, which is more than 7\% of the sample. A portion of these tweets include some sort of escalation indicator, such as hyperbole, sarcasm, or personal insults that critics tend to pick up on in their replies (see example in Table \ref{tab:mediacodes}. Once again, this definition of "escalation" was defined by those who criticised the tweets. In order to understand this further, we extended our analysis to look more deeply at the specific subject matter being discussed. The subjects Mr. Lammy is discussing are urgent and controversial in British political discourse, such as racism, Brexit and more personally, the leadership of Boris Johnson. It is not clear whether or not he is targeted because of his communication style (as critics say), his party membership, or because of his ethnic background. 

Affect and race are connected in how much anger and frustration those from a minority background are expected to express by a predominantly White society~\cite{ashley2014angry}. The stereotypes of the ``angry Black woman'' (or man) persist, in particular with connection to the topic of racism or inequality more generally. It has only been 33 years this past June that Black MPs have been elected to Parliament (https://labourlist.org/2020/06/watch-labour-celebrates-33rd-anniversary-of-first-black-mps-elected/). For this reason, it is worth considering how speech is understood through that lens.  



\section{Findings: Social Media Activities During COVID-19 (RQ3)}
\label{findings:socialmediaactivities}

Our third research question asked: \textit{Which social media activities of UK MPs during the COVID-19 pandemic receive the most abusive replies?}  
To answer this question, we applied our coding scheme described in \ref{subsec:thematic_analysis} to a sample of tweets that received a substantial number of replies that contained abusive language (see section \ref{intro}). The purpose of this was to identify qualitative differences in authors, content, or delivery that may help explain the negative discourse related to their tweets and highlight any other social factors at play. In total, we identified 190 tweets meeting these criteria. 

66 MPs authored the 190 tweets that received the highest number and percentages of abusive replies. 17 MPs are women, which is approximately 35\% of the sample. Women make up approximately 30\% of UK Parliament. \footnote{http://www.ukpolitical.info/female-members-of-parliament.htm} However, of the 17 women, 41\% are Women of Colour (n=7), though Women of Colour only make up a small percentage of an already small percentage MPs from a ``minority'' background. \footnote{https://theconversation.com/black-british-citizens-want-more-than-complacency-from-this-government-140297} It is important to note that none of these women are members of the Conservative party. In fact, Labour politicians have authored 108 of the 190 Tweets in this sample. Conservatives authored 53, Liberal Democrats 16, the Scottish National Party 11 and the Democratic Unionists 2.  To break this down further, we had 4 Conservative MPs who are female (all white), and 25 male MPs. For the Labour Party, that split is more even, with 11 women and 14 men. Table~\ref{tab:qualcorp} gives corpus statistics in terms of tweets authored. "Tweets" is number of tweets authored, "\% of corp" is the percentage of the qualitative corpus that number constitutes, and "\% Repr." is the representation that demographic has among MPs with Twitter accounts for comparison. "\# Replies" is the number of replies tweets by that demographic in the qualitative corpus received, and "\# Abusive" is the number of those replies that were abusive (recall that the tweet is only included if it receives a high level of abuse).

\begin{table}
\begin{tabular}{|l|rrrrr|}
\hline
\textbf{Demographic} & \textbf{Tweets} & \textbf{\% of corp} & \textbf{\% Repr.} &  \textbf{\# Replies} &  \textbf{\# Abusive}\\
\hline
White men & 348 & 72.63 & 60.63 & 257,960 & 28,263\\
Men of colour & 25 & 10.53 & 4.36 & 28,727 & 2,938\\
Women of colour & 32 & 7.89 & 5.57 & 30,999 & 3,198\\
White women & 169 & 8.95 & 29.44 & 33,435 & 3,391\\
\hline
\end{tabular}
\caption{Proportion of the qualitative corpus (most abused tweets) authored by different demographics, alongside representation of that demographic among MPs on Twitter for comparison, the number of replies and the number of abusive replies received by those tweets.}
\label{tab:qualcorp}
\end{table}

\subsection{Social Media Activities: Subjectivity in Escalation and Virtue}
Examining each tweet, we had 11 categories of social media activities, plus one additional ``unclear'' media category (see Table \ref{tab:mediacodes}). We added a modifier to the media activity of ``escalation'', if there were combinations of what we referred to as the five \textbf{indicators of escalation}: the presence of \textbf{hyperbole} (language that is perceived as having high valence), \textbf{sarcasm} or flippancy, \textbf{insult} and \textbf{abusive language}, making something \textbf{personal} (criticising the individual rather than their actions) or solidifying \textbf{``us and them''} narratives. These escalation indicators were derived from how critics and abusers of the tweets in our sample speak about escalating language and what they think is antagonistic. If a tweet only contained escalation indicators and no other content, it was categorised as ``escalation'' only. Examples of each category are provided in Table \ref{tab:mediacodes} and the full qualitative sample and coding notes are provided at the URL given in Section \ref{sec:declarations}. 

Escalations were particularly subjective and therefore difficult to classify. However, only 22 tweets contained some measure of escalation and 9 were coded as ``Escalation'' only. Here is an example of a tweet from Labour party member Ian Lavery that contains an escalation, in addition to its main media activity of rebuking authorities: \\
\url{https://twitter.com/IanLaveryMP/status/1238569895618625536}
\begin{quote}
    \textit{So does this “herd immunity” @BorisJohnson strategy mean accepting the end of life for many elderly \& vulnerable people But others should be fine ? Just asking for the elderly lady across the street.}
\end{quote}
This tweet was categorised as an escalation because it contains both sarcasm and hyperbole. 

Some tweets received criticisms and abuse related to a \textbf{particular event or pattern of behaviour}. We have 12 examples of this in our sample. Examples of this include Dominic Cummings' behaviour during Lockdown in May (see section \ref{subsub:May10}), the birth of Boris Johnson's child, or Sammy Wilson's previous voting record on the NHS (when combined with a tweet promoting clapping for the NHS). 

\begin{figure}
  \includegraphics[width=.95\textwidth]{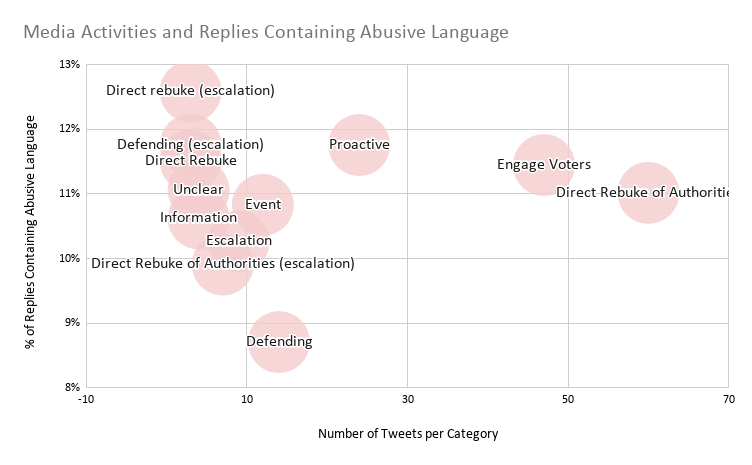}
  \caption{Media Activities and Replies Containing Abusive Language}
  \label{fig:SocMedia}
\end{figure}

Fig \ref{fig:SocMedia} shows the media activity category and the number of abusive replies per category. While tweets with escalations may attract the higher percentages of abusive replies, \textbf{the most common activities getting abusive replies are those ordinary to the job of an MP}. The most common media activities receiving replies that contain abusive language are \textbf{Direct Rebuke of Authorities} (n=60) and \textbf{Engaging Voters} (n=47). The following quote from Lisa Nandy (coded as a Direct Rebuke of Authorities) received 9\% replies including abusive language:

\url{https://twitter.com/lisanandy/status/1237339808017547264}

\begin{quote}
   \textit{It is irresponsible and short-sighted from the government to rule out extending the post-Brexit transition period. We should be taking action now to provide certainty for business in the face of this global economic challenge.}
\end{quote}

This is a fairly standard argument from a member of the opposition party who was a ``Remain'' voter, preferring a ``soft Brexit''.\footnote{https://www.theyworkforyou.com/mp/24831/lisa\_nandy/wigan/votes} Likewise, the following tweet from conservative Jack Lopresti is an attempt to speak to a core group voters and champion their interests: 

\url{https://twitter.com/JackLopresti/status/1248306181522763778}

\begin{quote}
    \textit{If off-licences and takeawayys are open, churches should be, Tory MP claims https://t.co/aA2CY06XrU}
\end{quote}

This tweet was sign-posting to an article in the Telegraph in which Mr. Lopresti makes his views known. This tweet attracted nearly 12\% abusive replies. 

Perhaps unsurprisingly, all of the parties receive replies that contain abusive language when they do the parts of their job that aggravate the other parties. For example, the parties in opposition criticise the party in power, which will defend itself. All parties attempt to reach voters. However, the left and left-centrist parties tend to reach out to voters that conservatives do not, such as religious minorities, migrants and People of Colour. The subject of ``virtue signalling'' arises in criticisms of this type of social media activity. This is evident in the large amount of abusive replies received by Liberal Democrats in response to their participation in Ramadan (see Section \ref{sec:events}).

As mentioned previously, virtue signalling is defined as communicating support for a specific issue with high moral value (such as fighting racism) without providing tangible support and effort~\cite{wallace2020consuming}. When the accusation of virtue signalling is levelled, the implicit assumption is that the gesture is empty or amounts to ``moral grandstanding''~\cite{tosi2016moral} without substance behind it. 
In this case, in the past 10 years, evidence shows that the Muslim community has shifted support from the Labour Party to the Liberal Democrats.\footnote{https://www.aljazeera.com/focus/britishelection/2010/05/20105312436485579.html} The party has responded to this, attempting repeatedly to elect a Muslim MP. The party has shown some attention to the social challenges of this group, suspending a candidate in 2019 for his comments online doubting the existence of Islamaphobia. \footnote{https://metro.co.uk/2019/04/18/lib-dem-candidate-suspended-comments-muslims-9243430/} They have also had a few gaffes, showing a lack of awareness of the culture \footnote{https://www.telegraph.co.uk/politics/2020/04/26/lib-dem-councillor-apologises-tweeting-photo-bacon-solidarity/} and have still not had a Muslim MP, despite efforts. \footnote{http://muslimnews.co.uk/newspaper/top-stories/record-18-muslim-mps-elected-majority-women/} Still, there is also evidence that the gesture of fasting and donating to charity as part of Ramadan was perceived as showing solidarity with the Muslim community during the COVID-19 pandemic.\footnote{https://www.easterneye.biz/lib-dem-mps-to-fast-during-ramadan-to-show-unity-for-muslim-community/} \textbf{Virtue signalling needs to be considered within a framework of whose attention is being courted and whether or not that community views the attention as tokenistic or meaningful}. 

\subsection{COVID-19 Topics: Priorities and Leadership}
\label{subsec:covidtopics}

Our second round of coding dealt with the COVID-19 subject referenced or alluded to in the tweet, which we determined inductively by going through the set of tweets using thematic analysis. We assigned tweets to one of 8 categories, including one ``non-COVID'' category (n = 37), if the topic was not related to COVID-19. These non-COVID topics include the floods that happened just prior to the pandemic, some more general thoughts on platform issues that are continuously relevant, such as budget and migration. Many non-COVID related tweets were posted before COVID-19 had reached the UK to such a significant extent. Topics include the cabinet reshuffle, class issues, Brexit, and migration. 

\begin{table}
\begin{tabular}
{|p{1.5cm}|p{1cm}|p{3cm}|p{5cm}|}
\hline
\textbf{COVID-19 Topic} & \textbf{No. Tweets} & \textbf{Description} & \textbf{Examples} \\
\hline
Covid \& Brexit & 10 & Discussing the impact of COVID-19 on Brexit & \allowbreak \textit{The country cannot afford the economic damage and chaos of a no-deal Brexit at the end of this year. Time is running out for the Government to agree to an extension and people’s livelihoods first.}- Layla Moran\\
\hline
Health Challenges \& Deaths (e) & 27 & Discussion of and reporting about fatalities during COVID-19 & \textit{Italy: Population - 60m Coronavirus cases - 21,157 Deaths - 1,441 UK: Population - 66m Coronavirus cases - 1,372 Deaths - 35 Italy had their first confirmed case 24hrs before the UK. The strategy is working.} - Rob Roberts\\
\hline
Finance \% Benefits & 23 & Discussion of how financial support and benefits will work during COVID-19. This includes discussion of what COVID-19 has shown us about our current economic models & \textit{There's no statutory sick pay for part-time, low-paid or zero-hours contract workers. And the rate of sick pay isn't enough to live on. Wrong at any time - but dangerous while people who might be ill are asked to stay home. The system is broken and now is the time to fix it.} - Jeremy Corbyn\\
\hline
Opposition Responses & 9 & Defense or rebuke of Labour MPs specifically & \textit{Intentionally misleading reporting is really disappointing at a time like this. I've spoken about the opportunity for people to get out there and help their local communities and those in need. Nonsense to suggest otherwise. We all need to do our bit to get through this crisis.} - Ian Lavery \\
\hline
Leadership \& Communication & 36 & Discussion of how policy or commentary has been delivered during the COVID-19 crisis & \textit{The @theSNP continue to let down thousands of homes in rural and remote communities. We need an audit on how they're spending broadband rollout money! See my exchange with the Government} - Jamie Stone\\
\hline
Lockdown \& Social Distancing & 28 & Discussion of how lockdown or social distancing is impacting people, the economy and the virus & \textit{I'm all in favour of wearing the appropriate face mask! @YesBikers} - Stewart Hosie\\
\hline
Minorities \& Undervalued Groups & 17 & Any discussion about a group that is viewed as a minority group in the UK & \textit{As senior Conservatives publish Sinophobic screeds in the rw press to distract from their own Government’s lethal complacency, it’s clear racism won’t stop for \#Coronavirus. Neither must we opposing it. Join the fightback in an hour! http://ow.ly/4GBm30qw1jX} - Bell Ribeiro-Addy\\
\hline
Non-COVID & 40 & Any discussion about topics that are not linked to Covid-19 in direct ways, such as flooding & \textit{A pre-dawn meal today Preparing for my first ever fast in the holy month of Ramadan For Muslims doing Ramadan in isolation, you are not alone! \#RamadanMubarak \#LibDemIftar} - Ed Davey \\
\hline
\textbf{Grand Total} & 190 &  &  \\
\hline

\end{tabular}

\caption{COVID-19 topics that are expressed MPs' posts on Twitter, to which the MP received abusive replies. Descriptions and examples included.}
\label{tab:covidcodes}
\end{table}

\begin{figure}
  \includegraphics[width=.95\textwidth]{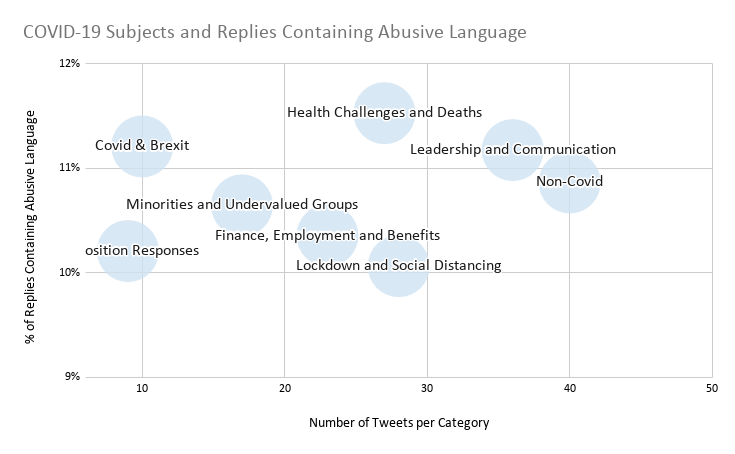}
  \caption{COVID-19  Subjects and Replies Containing Abusive Language}
  \label{fig:Covidtopics}
\end{figure}

After COVID-19 began to take hold, those non-COVID topics shift and are primarily related to specific issues or events, such as Jeremy Corbyn stepping down and Keir Starmer taking lead of the Labour Party, renewed references to a One UK policy/ One Parliament, Boris Johnson and the birth of his child, and the Liberal Democrats celebration of Ramadan. Within the COVID-19 topics, some more specific categories, such as ``fatalities" or the NHS were absorbed under the main category of Health challenges and deaths to arrive at groupings that including roughly the same number of examples from our data sample. The full list of categories can be found at the URL provided in Section \ref{sec:declarations}. In Fig \ref{fig:Covidtopics}, however, we show which topics around COVID-19 were associated with receiving more replies that are abusive in our qualitative sample. 

Leadership and Communication (n=39), along with Lockdown and Social Distancing issues (n=28) were the COVID-19 topics that had the best representation in the sample, with high number of replies that contain abusive language. These two categories include issues such as perceived government inaction and tone (Leadership and Communication), and guidance or impacts around lockdown or social distancing, such as wearing a face mask (Lockdown and Social Distancing). 

The following tweet from Labour MP Yvette Cooper attracted more than 8\% abusive replies, addressing perceived confusion around guidance from the UK government:

\url{https://twitter.com/YvetteCooperMP/status/1241849797155393537}

\begin{quote}
   \textit{I watched the Prime Minister’s press conference in despair. In a public health emergency communication and information saves lives. Yet time \& again the Government keeps failing to push out a strong clear message to everyone. For all our sakes they urgently need to get a grip.} 
\end{quote}

The following tweet from Jacob Rees-Mogg linked to an article about the Queen isolating herself amidst COVID-19 concerns. It attracted 11\% abusive replies, mostly for invoking the name of the Queen or comparing her experience to those of citizens who are struggling to meet their needs.

\url{https://twitter.com/Jacob_Rees_Mogg/status/1239949070426419200}

\begin{quote}
    \textit{As always an example to the nation: God save the Queen\\ https://t.co/N7egXeDzXD?amp=1}
\end{quote}

Discussion of Health Challenges and Deaths (n = 27) received the greatest percentage of abusive replies. This includes conversations around UK fatalities in comparison to other nations, support for the NHS and issues with testing. This type of comparison, as mentioned previously (especially from someone in a left-orientated party), is generally received as negative, and even unpatriotic in some critiques.














Women of Colour tended to discuss topics that address the needs of minorities and undervalued groups during COVID-19. White women have a profile more similar to men, in which questioning leadership and communication tended to be the COVID-19 subject for which they received more replies containing abusive language. Clearly, \textbf{questioning the government in power may lead to criticism}. Our literature review indicated that people tend to want to trust authorities during a crisis. 
Labour politicians' desire to keep the subject of racism and discrimination during COVID-19 at the forefront attracts some abusive comments for what is called \textbf{``playing politics", delivering ``low blows" or ``playing the race card"} to the party in power. Naturally, the opposition parties believe it is their job to rebuke authorities and suggest alternative policies. Likewise, racism is not seen as a platform issue, but a social issue that is continuously relevant. In this sense, \textbf{what is relevant to COVID-19 and should be prioritised} is being negotiated in some of this dialogue.

\subsection{Contentious Issues: Disrupting the Mainstream Discourse}
\label{subsec:contentiousissues}

\begin{figure}
  \includegraphics[width=.95\textwidth]{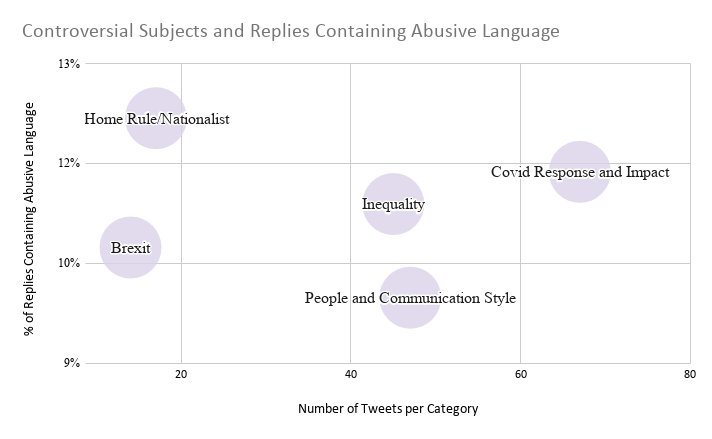}
  \caption{Contentious Issues and Replies Containing Abusive Language}
  \label{fig:controverytopics}
\end{figure}

\begin{figure}
  \includegraphics[width=.95\textwidth]{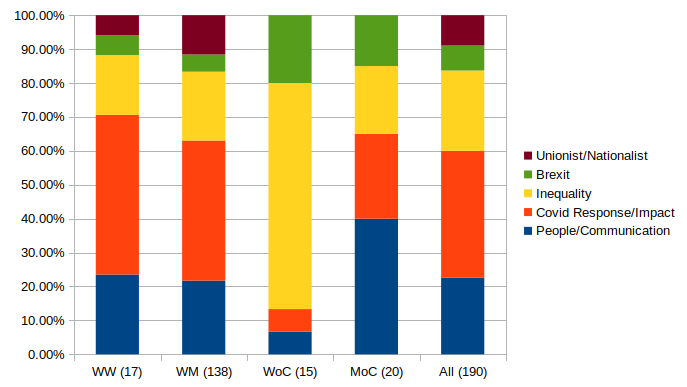}
  \caption{The 190 highly abused tweets in the qualitative sample, split by topic and by demographic. "WW" means white women, "WoC" means women of colour, and similarly for men.}
  \label{fig:touchy-subjects}
\end{figure}

When we looked deeper at the controversy that might be latent in the topics above, we identified more than 50 distinct subjects from our open coding, and had one category of ``unclear''. Some categories were related, such as Islam and Muslims, Racism and Immigration. We reduced the categories to 5 predominant issues (see Fig \ref{fig:controverytopics}): \textbf{Home Rule/Nationalist perspectives, Inequality and perceptions of inequality, Brexit} (a continued issue with new relevance), \textbf{COVID-19 Response and Impact}, and lastly, \textbf{People and Communication} (which includes subjects like personal folly, tone, etc.). Proportions of each subject to appear in tweets from different demographic groups, as well as overall, are shown in Fig~\ref{fig:touchy-subjects}. 

\begin{table}
\begin{tabular}
{|p{2cm}|p{1cm}|p{3cm}|p{4.5cm}|}
\hline
\textbf{Controversial Topic} & \textbf{No. Tweets} & \textbf{Description} & \textbf{Examples} \\
\hline
People \& Communication Style & 47 & Tweets about specific people and their actions or the way that they communicate. & \allowbreak \textit{Dom Cummings followed the guidelines and looked after his family. End of story.}- Oliver Dowden\\
\hline
Covid Response \& Impact & 67 & Tweets about the COVID-19 pandemic and any responses or impacts & \textit{Latest NHS advice: If you have: a new continuous cough OR a high temperature You should stay at home for seven days. Read the full guidance now: http://NHS.uk/coronavirus} - Boris Johnson\\
\hline
Inequality & 45 & Tweets about inequality in any form, be it racial, gender specific, religious, class-based, etc. & \textit{Amazing. Dave telling it like it is.} - John McDonnell\\
\hline
Brexit & 14 & Tweets about Brexit & \textit{Good to hear EU Commission saying they are fine with an Australia style deal or a bespoke Free Trade Agreement. So let's get on with it. No need to argue over it all year.} - John Redwood\\
\hline
Home Rule \& Nationalism & 17 & Tweets that appear to address long-standing conflicts about the Union and its Governance, including pro-Scotland and anti-SNP sentiment. & \textit{Really the most politically frustrating aspect of this crisis is not having an independent Scotland to do the emergency guaranteed income, to do the testing - to take a different path. We are instead shackled to inept economic extremists at Westminster.} - Angus MacNeil\\
\hline
\textbf{Grand Total} & 190 &  &  \\
\hline

\end{tabular}

\caption{Controversial topics that are expressed MPs' posts on Twitter, to which the MP received abusive replies. Descriptions and examples included.}
\label{tab:contentiouscodes}
\end{table}

For example, the \#stayalert slogan of the conservative government received considerable criticism for being confusing and potentially working against the goal of encouraging citizens to simply stay home. Criticising the government's efforts, MP Ian Blackford tweeted: 

\url{https://twitter.com/Ianblackford_MP/status/1259250871814295552}

\begin{displayquote}
\textit{``\#stayalert. What kind of buffoon thinks of this kind of nonsense. It is an invisible threat. Staying alert is not the answer \#StayHomeSaveLives is.''}
\end{displayquote}

This tweet received both a number of abusive replies to Mr. Blackford for critising the government's attempts to resolve complex challenges during the pandemic, and a number of supportive replies (some of which are also abusive toward the government and certain ministers). This polarisation indicates a liminal hotspot around trusting and critiquing authorities in crisis. 

In terms of party insights, the Liberal Democrats received a considerable amount of attention for speaking about Ramadan and the Muslim community (56\% of their sample). The SNP, as one might expect, gathers some abusive replies when tweeting support for Scottish independence or for promoting Scottish excellence. Hannah Bardell received abusive and critical replies for posting the following tweet:

\url{https://twitter.com/HannahB4LiviMP/status/1238750253182050310}
\begin{quote}

\textit{Once again the PM following Nicola Sturgeon’s lead}

\end{quote}

The tweet was accompanied by an article from the Guardian\footnote{https://twitter.com/guardiannews/status/1238591827449651205/photo/1} about the Prime Minister's decision to ban mass gatherings. 

Angus MacNeil was called ``divisive'' and ``divisionist'' for the following tweet in response to clapping for the NHS:

\begin{quote}
    \textit{It is "NHS Scotland"} 
\end{quote}

Likewise, making statements in favour of uniting the four UK countries, or making light of those countries' prerogative to handle the pandemic differently, attracts strong criticism and abuse. For example, Stephan Crabb received several abusive replies for the following tweet: 
\begin{quote}
    \textit{Quite the change in rhetoric from the days when Welsh Government were encouraging people to come to Wales and drive their 4x4's right onto the beaches.}
\end{quote}

Another liminal hotspot is around who is expected to speak about social issues of injustice, which commonly attract abusive responses. Women of Colour were speaking mostly about inequality when they received replies containing abuse (10 Tweets, which is 66\% of our sample of Women of Colour, and 5\% of our total sample). Men of Colour more closely resembled the subject attention of both White women and White men on the more general discussion of COVID-19 response and leadership through the crisis (For all men, issues with Leadership, or COVID-19 response made up 60\% of all topics; for white women, more than 70\%. 10\% of the topics discussed by men in their sample are about controversial people specifically, such as Boris Johnson, Jeremy Corbyn and Dominic Cummings).

\section{Discussion and Future Work} 
All activity online can be viewed as communication and persuasion - there are people on different sides of different issues, vying for the public attention. This can attract positive and negative responses. However, overall, what our investigation has shown us is that dimensions of political discourse are mediated by perceptions of power, potentially due to the uncertain situation created by COVID-19. In the sections below, we summarise our findings about the influence of ideology, political authority and affect on how the words of MPs are communicated and interpreted by the public during the COVID-19 pandemic. 

\subsubsection{Power, Ideology and Virtue}
When parties on the left speak directly with and from undervalued communities or voters, this may be perceived as virtue signalling by the right. When a party is not in power, this is even more difficult to communicate. However, when the party in power shows a lack of tact toward excluded communities, this is also judged more harshly.

When the party in power communicates policy about controversial issues in the name of the people, it will get push-back from people who did not vote for that party. When the left attempts to meet voters by discussing issues (racism, migration) that a portion of the public may feel are external to relevant British politics right now (COVID-19 and Brexit), they will get abuse. However, for the left, those are topics that exist everywhere all the time and are systemic. Looking at successful interventions by opposition parties in the government's activities may constitute an interesting area for future research. More specifically, this research could help to answer questions about the origins of priorities and disagreements.  

\subsubsection{Power, Authority and Vindication}
What the answers to our three research questions indicate is that it matters \textbf{who is ``in charge''}, when looking at how the public and other ministers respond to the social media activities of UK MPs during COVID-19. The party in power (along with its members) will have more responsibility to the public for mastering tone and explaining their actions. Opposition parties will have more difficulty in a health crisis to not be perceived as unnecessarily antagonistic. 
In addition, we found that it matters who a person is and what they represent, \textbf{whether or not an individual will be perceived as a trustworthy authority}. Many tweets in our sample appeared to be speaking to core groups of voters and other parliamentarians. They are not necessarily an invitation for debate. 

\subsubsection{Power, Affect and Vitriol}
In terms of affect, our data shows that it matters what you say and how you say it, particularly \textbf{in connection with priors}. If Jeremy Corbyn posts about racism, and has been continuously in the news for not handling antisemitism in his party, he will get some angry replies. If Sammy Wilson voted against a pay-rise for nurses in 2017, and then posts a ``clap for carers'' post on Twitter, those who remember his prior voting record will be angry. The more affected a tweet is, the more this appears to aggravate.

Vitriol as a result of one's previous political statements or actions is one side of the story. Hate is another. The issue of \textbf{what is abusive \textit{versus} what is hate speech} needs to be disentangled from both abuse and from racism. Racism does not only involve hate speech. It also involves a) expecting People of Colour to champion racial equality, as the breakdown of topics indicates and b) framing racism as a fringe issue. This is evidenced in our dataset. Abuse, though uncomfortable and uncivil, is a different type of speech whose study may be useful for any number of discussions, potentially on the subject of agonism or counter-speech. Agonism argues that the contestations of the time can be used to renew democracy and strengthen public discourse~\cite{machin2019democracy}. Promising work on recognising an highlighting counter-speech in online communication is already on the horizon~\cite{chung2019conan,mathew2019thou}. 

\section{Conclusion}
In this paper, we explored how UK MPs contribute  to  the  information  and  communication  environment during COVID-19, and the abusive replies that they receive. Contextualising these activities in terms of what the public expect during a health crisis, how ministers typically use social media to communicate in crisis, and which mitigating features of either the person or context interfere in those activities, we were able to advance the conversation about online abuse toward some new directions, such as how to understand virtue signalling or what it means to play party politics. Building on previous studies of  abusive  language  toward  British  MPs,  we  offered  a  large-scale, mixed-methods study of abusive and antagonistic responses to UK politicians during the COVID-19 pandemic from early February to late May 2020. We found that - similarly to other key moments in British contemporary politics - political ideology, authority and affect have played a role in how MPs social media posts were received by the public. In the context of COVID-19, we found that pressing  subjects, like financial support or unemployment, may attract high levels of engagement, but do not necessarily lead to abusive dialogue. As with earlier findings, prominence and event surges impact the amount of abusive replies MPs received. In addition, the topic of the tweet (in particular if it is divisive) and the individual bringing that topic into discussion (their gender, ethnicity or party, for example) impacted levels of abuse. Women of Colour appear to bring the topic of inequality to the table and this attracts a variety of abuse. Other MPs may be discussing inequality and not receiving abuse (which this work did not cover).

In conclusion, this work contributes to the wider understanding of abusive language online, in particular that which is directed at public officials. Issues of power, which are crystallised in terms of political power or social power impact communication at all stages. 

\section*{Declarations}
\label{sec:declarations}
\subsubsection*{Funding:}
This work was supported by the ESRC under grant number ES/T012714/1, ``Responsible AI for Inclusive, Democratic Societies: A cross-disciplinary approach to detecting and countering abusive language online.''

\subsubsection*{Conflicts of interest/Competing interests:} 
No conflicts of interest or competing interests.

\subsubsection*{Availability of data and material:} 
Data is available here:\\

\url{https://gate-socmedia.group.shef.ac.uk/abuse-mps-covid/}

\subsubsection*{Code availability:} 
Not applicable.

\begin{acknowledgements}
Thanks to Mehmet E. Bakir for comments and suggestions.
\end{acknowledgements}

%
%

\bibliographystyle{spmpsci}      
\bibliography{COVID.bib}   

\end{document}